\documentclass[fleqn,12pt]{article}
\usepackage{latexsym,amsmath,tabularx,amssymb,bm}
\usepackage{epsfig}
\newcommand{\nn}{\nonumber\\ }
\newcommand{\beq}{\begin{eqnarray}}
\newcommand{\eeq}{\end{eqnarray}}
\def\simge{\mathrel{%
   \rlap{\raise 0.511ex \hbox{$>$}}{\lower 0.511ex \hbox{$\sim$}}}}
\def\simle{\mathrel{
   \rlap{\raise 0.511ex \hbox{$<$}}{\lower 0.511ex \hbox{$\sim$}}}}

\begin{document}
\begin{flushright} 
SACLAY--T03/138\\
{CU-TP-1096}
\end{flushright}
\vskip50pt
\begin{center}
\begin{title}
\title{\Large\bf Rare fluctuations and the high--energy limit
\vskip 6pt
of the $S$--matrix in QCD}
\vskip 10pt\vskip 10pt
{{\large Edmond Iancu}\\
{\it Service de Physique Th\'eorique, CEA Saclay\\
91191 Gif-sur-Yvette cedex, France}
\vskip 10pt
and
\vskip 10pt
{\large A.H. Mueller\footnote{This research is supported 
in part by the US Department of
Energy.}}\\
{\it Department of Physics, Columbia University\\
New York, New York 10027, USA}}
\end{title}

\vskip 50pt
{\bf Abstract}
\end{center}
\vskip8pt

We argue that one cannot correctly calculate the elastic
scattering $S$--matrix for high-energy dipole-dipole scattering, in the
region where $S$ is small, without taking fluctuations into account.
The relevant fluctuations are rare and unimportant for general
properties of inelastic collisions.  We find that the Kovchegov
equation, while giving the form of the $S$--matrix correctly, gives the
exponential factor twice as large as the result which emerges when
fluctuations are taken into account.

%
\newpage
\section{Introduction}

 The topic of this paper is understanding the role that
fluctuations play when unitarity corrections become important in
high-energy scattering.  We focus our attention on dipole-dipole
scattering because the issues are a bit sharper there.  We can perhaps
motivate our discussion by the following observations.  At very high
energy the typical (mean) configuration of a dipole's light-cone
wavefunction is a Color Glass Condensate [1--4], a state characterized
by a saturation \cite{bov,ler} momentum, $Q_s,$ and having high
occupancy for all gluonic levels of momentum less than or equal to
$Q_s,$ which depends on the rapidity, $y,$ of the parent dipole.
Suppose we scatter two such evolved dipoles at zero impact parameter,
in the center of mass frame and at relative rapidity $Y.$ Then, if we use
just these typical configurations, of the condensate type, to compute
the $S$--matrix for the scattering, the result will be proportional to
$\exp\{-const \,Q_s^2(Y/2) r_0^2/\alpha_s^2\},$ where $r_0$ is the
size of the parent dipoles serving as the seeds for Color Glass
Condensates which collide.  The $Q_s^2r_0^2$ part of this formula is
purely geometric; it is the number of, roughly, independent parts of
one of the condensates when viewed on a scale $\Delta x_\perp \sim
1/Q_s.$ A more complete discussion of condensate-condensate scattering
is given in section 3.3 with the $S$--matrix given in eq. (25).  On the other
hand the Kovchegov equation \cite{Kov} gives an $S$--matrix (see eq. (14))
which is much larger than (25), being proportional to 
$\exp\{-const\,\ln^2(Q_s^2r_0^2)\}.$ Why do eqs. (14) and (25) disagree and which of
these answers is correct ?  

 The problem is in the use of typical, or mean, configurations
to estimate the elastic scattering $S$--matrix.  If one collides in the
center of mass frame two Color Glass Condensates, then the exponent we
arrived at, $Q_s^2r_0^2/\alpha_s^2$, corresponds to the average number
of gluons produced in the collision.  The condensate description of
the wavefunction is a good starting point for a typical (inelastic)
collision, but it need not be correct for evaluating a very small
observable like the elastic $S$--matrix whose evaluation can be (and is)
dominated by much rarer configurations.  As we shall see the parts of
the wavefunction which dominate the $S$--matrix are very dependent on the
frame in which we view the scattering.  In the center of mass frame,
and in the high-energy regime where $S$ is very small, the $S$--matrix
is dominated by wavefunction configurations which have relatively few
gluons \cite{uel} and which are not in a condensate state.  In an
asymmetric frame, the parent dipole having the lower momentum evolves
into a state with few gluons while the higher momentum dipole evolves
into a condensate, but into a condensate having a lower than normal
saturation momentum.  

 What about the result (14) coming from the Kovchegov equation?
While the functional form of the solution \cite{ncu,vin} is fine,
 we find that the
Kovchegov equation misses the correct evaluation of the exponent of
$S$ by a factor of two.  Again the problem is that the Kovchegov
equation does not treat fluctuations properly. The crucial ingredient
in deriving the Kovchegov equation is the replacement of
$S_Y^{(2)}(x_\perp -z_\perp, z_\perp -y_\perp),$ the scattering
$S-$matrix for a two-dipole state (dipoles of sizes $x_\perp -
z_\perp$ \ and\ $z_\perp - y_\perp)$ by the product
$S_Y(x_\perp-z_\perp)S_Y(z_\perp-y_\perp).$ Such a replacement is true
only in the absence of fluctuations in the light-cone wavefunction
of the target.
Contrary to a rather widely held belief this replacement is not
justified by large $N_c$ behavior since, as we shall see below, the
relevant fluctuations concern a suppression of gluons in certain
momentum regions of the wavefunction, an issue which is
independent of $N_c$.  Despite these problems, the Kovchegov equation is
probably the best one can do with respect to a simple equation which
naturally imposes unitarity on BFKL evolution.

  Much of what we say here has been anticipated in
some early works \cite{uel,ala} on the study of unitarity in
dipole-dipole scattering.  The issue of fluctuations was reasonably
well understood although at that time it was not possible to achieve
the level of analytical precision that we are able to present here.
The detailed numerical studies done by Salam \cite{ala} correspond to a
(numerical) evaluation of the Balitsky equation \cite{lit} or,
equivalently, of the functional evolution equation for the Color Glass
Condensate \cite{anc,Ian,vne,ert}, rather than an evaluation of the
Kovchegov equation.

 Finally, it must be admitted that we are not able to prove
that the wavefunction configurations which we suggest are dominant are
in fact so.  Certainly, the configurations that we shall find are better
than those chosen by the Kovchegov equation since our $S$--matrix is
much bigger.  And we do give arguments in sections 3.6 and 4 to the end
that natural modifications of our choice of configurations lead to a
smaller $S$--matrix.  Nevertheless, it is difficult to classify all the
possisble configurations so as to be sure that we have made the right
choice.  In this context it should be noted that the coefficient of
the $(\bar{\alpha}_sY)^2$ term in the exponent of our result (31) is
about a factor of 1.7 higher than the numerical evaluation in Ref.
\cite{uel}.
We do not understand the resolution of the discrepancy.

 In Sec. 2 we give a rapid derivation of the Kovchegov equation,
and of its prediction for the high-energy behavior of the 
$S$-matrix for dipole-dipole scattering in the regime where $S$ is
very small. The usual Levin--Tuchin result \cite{ncu,vin}
is recovered.  

In Sec. 3, we investigate dipole-dipole
scattering at very high energies in the center of mass (CM)
frame and at fixed impact parameter.  In Sec. 3.1 we
review the BFKL-dipole picture.  In Sec. 3.2, we show that using BFKL
evolution for the dipole wavefunctions, and taking into account only
average properties like their respective dipole number densities,
leads to an $S$-matrix which is too small, even though we work in a
rapidity region where BFKL evolution is correct for average properties
of the wavefunction.  In Sec. 3.3 we collide two Color Glass Condenses,
and again find an $S$-matrix which is unrealistically small. In
Sec. 3.4 we describe which (rare) confingurations lead to the result of
the Kovchegov equation in a natural asymmetric frame.  In Sec. 3.5 we
describe a set of rare fluctuations which give a much larger
$S$--matrix and which we suggest dominate very high-energy scattering
in the CM frame.  In Sec. 3.6 we give (incomplete) arguments as to why
the configurations we have chosen are optimal.

 In Sec. 4, we transform our 
center of mass picture to an arbitrary frame 
and describe the picture which emerges there.

\section{The Kovchegov equation}

  The Kovchegov equation \cite{Kov} is probably the
best ``simple'' equation for dealing with the onset of unitarity in
high energy, but weak coupling, scattering in QCD.  The argument for
the Kovchegov equation is easily stated. Consider the high-energy
scattering of a dipole, consisting of a quark at $x_\perp$ and an
antiquark at $y_\perp,$ on a target which may be another dipole or a
more complex hadron or nucleus.  It is convenient to view the
scattering in a frame where the dipole is going along the negative
$z$-axis (left-moving) and the target is going along the positive
$z$-axis (right-moving), and where almost all of the rapidity, $Y,$ of
the scattering is taken up by the right-moving system. 
(This is sometimes referred to as the ``dipole frame''.)  In this frame
the process takes the form of an elementary (unevolved) dipole of size
$x_\perp-y_\perp$ scattering in the field of a (highly evolved)
target.  If $S_Y(x_\perp,y_\perp)$ is the $S$--matrix for the
scattering we define the amplitude ${\cal N}_{xy}$ by
\begin{equation}
{\cal N}_{xy}=1-S_Y(x_\perp,y_\perp).
\end{equation}
 Now suppose we increase $Y$ by a small amount.  If this
increase is given to the target, then in order to calculate the change
in $S_Y$, it is necessary to calculate how the wavefunction of the
target changes with $Y.$ This change can be calculated using a functional
equation derived by Jalilian-Marian, Iancu, McLerran, Weigert,
Leonidov and Kovner (JIMWLK) \cite{anc,Ian,vne,ert}.   However, in order
to derive the Kovchegov equation it is simpler to keep the rapidity of
the target fixed and put the small change of rapidity into the
left-moving elementary dipole.  The dipole now has a small probability
of emitting a gluon due to this change of rapidity.  If the gluon is
in the wavefunction of the dipole at the time it scatters on the
target, then what scatters is a quark-antiquark-gluon
system which, in the large $N_c$ limit, can be viewed as a system of
two dipoles.  (One of these dipoles is the original quark and the
antiquark part of the gluon while the other dipole is the quark part
of the gluon and the original antiquark.)  If the gluon is not in the
wavefunction at the time of the scattering, it can be viewed as the
``virtual'' term which decreases the probability that the
quark-antiquark pair remain a simple dipole, compensating the
probability for the two--dipole state.  One can put this in formulae as
\beq
{\partial\over \partial y}\,S_Y(x_\perp,y_\perp)&=&{\bar{\alpha}_s\over 2\pi}\int
d^2z_\perp{(x_\perp-y_\perp)^2\over
(x_\perp-z_\perp)^2(z_\perp-y_\perp)^2}\nn
&{}&\quad\times
\left[S_Y^{(2)}(x_\perp,z_\perp,y_\perp)-S_Y(x_\perp,y_\perp)\right],
\eeq
 where $\bar{\alpha}_s={\alpha}_sN_c/\pi$,
$z_\perp$ is the transverse coordinate of the emitted
gluon, and $S^{(2)}_Y$ stands for the scattering of the two dipole
left-moving system on the target.

 If the target is homogeneous on a scale large compared to
$x_\perp-y_\perp$, then it is convenient to view $S_Y$ as a function of
$x_\perp-y_\perp$ and $b_\perp={x_\perp + y_\perp\over 2}$, and
$S_Y^{(2)}$ as a function of $x_\perp-y_\perp, z_\perp-y_\perp$ and
$b_\perp.$ Suppressing the $b_\perp$--dependence, eq. (2) becomes
\beq
{\partial\over \partial y} \,S_Y(x_\perp-y_\perp)&=&
 {\bar{\alpha}_s\over 2\pi} \int d^2z_\perp{(x_\perp-y_\perp)^2\over 
(x_\perp-z_\perp)^2(z_\perp-y_\perp)^2}\nn &{}&\quad\times
\left[S^{(2)}_Y(x_\perp-z_\perp,z_\perp-y_\perp) - 
S_Y(x_\perp-y_\perp)\right].
\eeq
 This is not yet the Kovchegov equation, rather it is part of
a set of equations derived by Balitsky \cite{lit}, and which also follows
from the functional evolution equation in Refs. \cite{anc,iro,ert}.
Eq. (3) is very
difficult to use because a solution for the $Y$--dependence of $S_Y$
requires knowing $S_Y^{(2)}.$ One obtains the Kovchegov equation if
one further assumes
\begin{equation}
S_Y^{(2)}(x_\perp-z_\perp,z_\perp-y_\perp) = S_Y(x_\perp-z_\perp)
S_Y(z_\perp-y_\perp),
\end{equation}
which is a sort of mean field approximation for the gluonic fields in
the target.  In his original papers \cite{Kov},  Kovchegov has obtained eq. (4)
for the case that the target is a large nucleus; in that case, this equation
should be a good approximation so long as $Y$ is not so large as to
cause $S$ to be dominated by nuclear density fluctuations.  However
for general targets one cannot expect (4) to be exact, as we shall see
in the next section.  Nevertheless, (4) may be a reasonable
approximation as unitarity corrections are just becoming important,
and in any case it leads to an interesting equation.

Specifically, using the approximation (4) in eq.~(3) gives
\beq
{\partial\over \partial y} \,S_Y(x_\perp-y_\perp) &=& {\bar{\alpha}_s\over 2\pi} \int
d^2z_\perp {(x_\perp-y_\perp)^2\over
(x_\perp-z_\perp)^2(z_\perp-y_\perp)^2}\\ &{}&\quad\times
\big[S_Y(x_\perp-z_\perp)S_Y(z_\perp-y_\perp)-S_Y(x_\perp-y_\perp)\big],
\nonumber\eeq
 or, equivalently,
\beq
{\partial\over \partial y}\, {\cal N}_{xy}&=&{\bar{\alpha}_s\over 2\pi} \int d^2z_\perp
{(x_\perp-y_\perp)^2\over (x_\perp-z_\perp)^2(z_\perp-y_\perp)^2}
\nn &{}&\qquad\quad\times\big[{\cal N}_{xz}+ {\cal N}_{zy}-{\cal N}_{xy}-{\cal N}_{xz}\,{\cal N}_{zy}\big],
\eeq
 which are two common forms of the Kovchegov equation \cite{Kov}.
Eq. (6) is the more useful equation when scattering is weak.  In that
case the quadratic term in (6) may be dropped and the dipole version
of the BFKL equation results.  Eq. (5) is easier to use when $S$ is
small, and unitarity corrections have become very important.

But in the general case,  when neither $S$ nor ${\cal N}$ is small,
eqs. (5) or (6) are difficult to deal with analytically.  There
are a number of good numerical evaluations [14--19] of the Kovchegov
equation which do cover the region where $S$ is neither close to zero
nor close to one. Here, however, we are interested only in the high-energy
regime where unitarity corrections are very important, so $S$ is small
indeed, and the solution to eq. (5) can be evaluated
analytically.  If $S_Y(x_\perp-y_\perp)$ is very small, then the
quadratic term in $S$ on the right-hand side of (5) is much smaller
than the linear term unless either $(x_\perp-z_\perp)^2$ or
$(z_\perp-y_\perp)^2$ is as small as $Q_s^{-2}(Y).$ We have introduced the
saturation momentum, $Q_s(Y),$ which is an intrinsic scale of the target
(characteristic of the gluon density there), and which
marks the scale at which a dipole scattering off the target
makes the transition from weak ($r_\perp\ll 1/Q_s$) 
to strong ($r_\perp\gg 1/Q_s$) interactions ($r_\perp$ is the
size of the dipole). More precisely, it is common to define
$Q_s(Y)$ by the equation $S_Y(r_\perp)=1/e$ when $r_\perp=2/Q_s.$

The regime of interest here corresponds to $r_0 \equiv 
\vert\,x_\perp-y_\perp\vert\gg 2/Q_s$.
Then, the right-hand side of (5) is dominated by the
logarithmic regions of integration where one has either
\begin{equation}
4/Q_s^2 \ll (x_\perp-z_\perp)^2 \ll r_0^2,
\end{equation}
 or
\begin{equation}
4/Q_s^2 \ll (z_\perp-y_\perp)^2 \ll r_0^2.
\end{equation}
 One easily finds
\begin{equation}
{\partial\over \partial y} \ln S_Y(r_0) \,\simeq\, -
\bar{\alpha}_s\int_{Q_s^{-2}(Y)}^{r_0^2}{dr^2\over r^2}\, =\,
- {\bar{\alpha}_s} \ln[Q_s^2(Y)r^2_0].
\end{equation}
 Now, we know that (see, e.g., \cite{Ian})
\begin{equation}
\ln[Q_s^2(Y)r^2_0]=c\bar{\alpha}_s(Y-Y_0) + \cdot\cdot\cdot
\end{equation}
  where $Y_0$ is such that $Q_s(Y_0) \sim 1/r_0$, and \cite{bov,AM99}
\begin{equation}
c \,\equiv \,{2\chi(\lambda_0)\over 1-\lambda_0} \simeq 4.883\,.
\end{equation}
 In the equation above,  $\lambda_0$ is the solution to
\begin{equation}
\chi(\lambda_0) = - \chi^\prime(\lambda_0)(1-\lambda_0),
\end{equation}
 where
\begin{equation}
\chi(\lambda) =\psi(1) - {1\over 2} \psi(\lambda) - {1\over 2} \psi(1-\lambda)
\end{equation}
 is the usual BFKL \cite{aev,Bal} eigenvalue function.  The
omitted terms on the right-hand side of (10) have only a logarithmic
dependence on $Y.$ Using (10) in (9) one finds \cite{ncu,Ian,vin}
\begin{equation}
S_Y(r_0) = {\rm e}^{-{c\over 2}\bar{\alpha}_s^2(Y-Y_0)^2}S_{Y_0}(r_0)\,,
\end{equation}
where $S_{Y_0}(r_0)\sim 1$.
 
Because we
have not kept $\ln(Y-Y_0)$ terms on the right-hand side of (10),
there are some missing linear terms in $Y$ in the exponent of eq.~(14).
Our focus in this paper will be primarily on the quadratic term, and
its coefficient, in the exponent of (14).

 Eq. (14) gives the standard result in the literature.  We have
gone through such a detailed ``derivation'' of (14) because the main
point of the present paper is to argue that (14) is in fact not quite
right.  The problem, as we shall see, is not with the derivation of
(14) from (5), but with the mean field approximation contained in (4).
Deep in the saturation region, where $S$ is very small, fluctuations
away from average configurations are actually dominant for the calculation
of $S$.

\section{Dipole-dipole scattering in the CM system}

 In this section we consider dipole-dipole scattering in the
center of mass frame.  Of course, the scattering cross section
cannot depend on the frame in which it is calculatd, however, the form
that the calculation takes can be very different in different frames
in a non-covariant gauge calculation.  In the first three parts of
this section we proceed in a qualitative manner, in order that the
physics issues not be obscured by technical details.  In the final
three parts of this section we proceed more technically and arrive at
an evaluation of the $S$--matrix which disagrees with (14), being a
factor of two smaller in the exponent.

\subsection{The BFKL approximation}

 Suppose we scatter two dipoles, each of size $r_0,$ at a
relative rapidity $Y.$ Then in a frame where one of the dipoles has
rapidity $y$ and the other has rapidity $Y-y$ the BFKL approximation
gives \cite{lle,tel}
\begin{equation}
\sigma(Y)=\int{d^2r_1d^2r_2\over
4\pi^2r_1^2r_2^2}n(r_0,r_1,y)n(r_0,r_2,Y-y)\sigma^{(0)}(r_1,r_2)
\end{equation}
 where $n(r_0,r_1,y)$ is the number density of radiated dipoles of
size $r_1$ in the wavefunction of a parent dipole of size $r_0$,
within a rapidity interval equal to $y$. 
Furthermore, $\sigma^{(0)}$ is the lowest order dipole--dipole
scattering cross section,  given by
\begin{equation}
\sigma^{(0)}(r_1,r_2) = 2\pi \alpha_s^2 r_<^2
\left(1+\ln\frac{ r_>}{r_<}\right),
\end{equation}
 where $r_<$ is the smaller of $r_1,r_2$ and $r_>$ is the
larger of $r_1,r_2.$ In the saddle point 
approximation to the BFKL solution,
\begin{equation}
n(r_0,r_1,y)\, \approx\, {r_0\over 2r_1}\,{{\rm e}^{(\alpha_P-1)y}\over {\sqrt{{7\over
2}\alpha_sN_c\zeta(3)y}}}\,\exp\left\{-{\ln ^2(r_0/r_1)\over
14\bar{\alpha}_s\zeta(3)y}\right\}
\end{equation}
 leading to
\begin{equation}
\sigma(Y) \,\approx\, 4\pi\alpha_s^2 r_0^2 \,
{{\rm e}^{(\alpha_P-1)Y}\over {\sqrt{{7\over
2}\alpha_sN_c\zeta(3)Y}}}.
\end{equation}
In these equations, $\alpha_P-1= 4\bar\alpha_s\ln 2$ is the BFKL `intercept'.
 What we are interested in here is not so much the precise
formulae contained in eqs. (15)--(18), but rather the qualitative picture.
To that end we rewrite the cross-section, eq. (18), as
\begin{equation}
\sigma(Y) = \pi r_0^2\alpha_s^2 n^2(r_0,r_0,Y/2) C(\alpha_sY)
\end{equation}
 where
\begin{equation}
C(\alpha_sY) = {8 {\sqrt{{7\over 2}\alpha_sN_c\zeta(3)Y}}}
\end{equation}
is an uninteresting and slowly varying prefactor. Eq. (19)
gives the cross section in terms of an elementary dipole--dipole cross
section, $\alpha_s^2\pi r_0^2,$ times the number of evolved dipoles in
the wavefunctions of each of the colliding parent dipoles.  Eq. (19) is
a reliable formula so long as $\sigma$ is well below $\pi r_0^2.$ When
$\sigma$ approaches $\pi r_0^2$, unitarity corrections become important
and it is to this topic that we next turn.

\subsection{The onset of unitarity corrections}

 Let $Y_0$ be the rapidity where $\sigma(Y_0)$ is equal to $\pi r_0^2.$  
Then, roughly,
\begin{equation}
Y_0 \simeq {1\over \alpha_P-1} \ln \frac{1}{\alpha_s^2}\,,
\end{equation}
 where the neglected terms in (21) are of size ${1\over
\alpha_P-1} \ln \ln 1/\alpha_s.$ Eq. (17) gives the average
number of evolved dipoles in the parent dipole wavefunction so long as
$y < Y_0$, at which point saturation effects in the dipole
wavefunction become important.  (The fact that the wavefunction of a
dipole does not receive significant saturation corrections
in the region $Y\ll Y_0$ \cite{Mue} follows from eq.~(15), which if we take
$y$ small requires that the result (18) come completely from
$n(r_0,r_2,Y).$ This also shows that, when the scattering is seen in
an {\it asymmetric} frame, unitarity corrections and saturation effects in 
the wavefunction of the evolved dipole start to be important at the
same rapidity\footnote{This is in agreement with the discussion
in Sec. 2, where we have seen that, in the dipole frame, the critical
rapidity for strong scattering $Y_0$ is related to the saturation scale:
$Q_s(Y_0) \sim 1/r_0$, cf. eq.~(10).}, namely at $Y=Y_0$.)

  Now consider dipole--dipole scattering in the CM
frame with the two parent dipoles scattering at zero impact parameter
and at rapidity $Y \lesssim 2Y_0.$ What is the $S$-matrix for such a
collision?  Of course when $Y > Y_0$ we expect very strong unitarity
corrections to the scattering. However, it should be reasonable, so
long as $Y < 2Y_0,$ to assume that the individual wavefunctions of
the colliding dipoles are still given by BFKL evolution (since 
$y=Y-y=Y/2 < Y_0$).  Now $S_Y^2$ has the
interpretation of being the probability that no interaction take place
in the collision. It is tempting to estimate 
$S^2_Y$ by requiring that no evolved
dipole in one wavefunction interact with any dipole in the other
wavefunction.  This gives
\begin{equation}
\ln S_Y^2(r_0) \,\simeq \,-  \,c_0\,\alpha_s^2n^2(r_0,r_0,Y/2) 
\end{equation}
 where the constant factor $c_0$
is not under control.  When $Y \simeq
2Y_0,$ using (17) and (21) in (22) gives
\begin{equation}
S_{2Y_0}(r_0) \simeq {\rm e}^{-c/\alpha_s^2}
\end{equation}
 which is much smaller than what one would obtain using the
result (14) from the Kovchegov equation along with the estimate (21).

Eq. (23) is clearly not right. The reason for such a failure is
the fact that the above calculation of the $S$-matrix in the presence
of unitarity corrections, cf. eqs. (22)--(23), has included 
only {\it typical} configurations in the wavefunctions of the parent
dipoles. Of course, the use of typical configurations is correct
as long as the energy is not too high ($Y < Y_0$), so that
one can restrict oneself to the single scattering 
(or ``single pomeron exchange'') approximation, eq. (18).
But at higher energies ($Y>Y_0$), where multiple collisions {\it are}
important, the wavefunction correlations among the dipoles which
scatter simultaneously play a crucial role, and lead to a result for
$S$ which is very different from eqs. (22)--(23). 
So long as $Y < 2Y_0$, and for center of mass scattering, the relevant
correlations are correctly described by BFKL evolution, but this has to be
applied to the detailed dynamics of all the dipole configurations, 
and not only to average properties, like the mean dipole number density (17).

The correct calculation which sums up ``multiple pomeron exchanges''
and replaces eqs. (22)--(23) in the regime where $Y_0 < Y  < 2Y_0$
is actually known. This has been originally developed within the color
dipole picture \cite{lle,tel,Mue}, and recently rederived within the
color glass formalism \cite{IM03}. This is a rather complex calculation
which cannot be analytically completed, but has been implemented numerically
by Salam \cite{ala}. As mentioned in the Introduction, the numerical
results thus obtained \cite{uel} are consistent with the functional form
in eq. (14), but with a smaller coefficient in the exponent.
As we shall see later in this section, such a smaller coefficient
emerges indeed when some rare configurations in the wavefunction
are taken into account.

\subsection{The collision of two Color Glass Condensates}

  In order to emphasize that typical wavefunction
configurations can give very wrong answers when $S$ is small, we now
estimate, from typical configurations, the $S$--matrix when $Y \gg
Y_0.$ Then, in the CM frame, what collide are two Color Glass
Condensates, each characterized by a saturation momentum $Q_s(Y/2).$
The number of gluons having $k_\perp \sim Q_s$ in each wavefunction is
\begin{equation}
N\simeq c_1\,{1\over \alpha_s^2}\, Q_s^2(Y/2) r_0^2
\end{equation}
 where, as before, $r_0$ is the size of the parent dipole in
each of the colliding condensates.  $c_1$ is a constant which is not
important for our purposes.  Again interpreting $S^2_Y$ as the
probability that no collision takes place one easily gets
\begin{equation}
S_Y\simeq \exp\left\{-c_2\,{1\over \alpha_s^2}\,Q_s^2(Y/2) r_0^2\right\}
\end{equation}
where the factor $Q_s^2r_0^2$ in (24) and (25) counts the
number of independent regions availabe for gluons, with $k_\perp \sim
1/Q_s,$ to occupy in an area $\pi r_0^2.$  Alternatively,
the result in eq. (25) could be obtained by using the formalism developed
in Ref. \cite{IM03} for the scattering between two color glasses, 
and focusing on the typical configurations,
of the condensate type, in both dipoles.

When $Y\rightarrow 2Y_0$,
$Q_s^2(Y/2) \rightarrow Q_s^2(Y_0) \simeq 1/r_0^2$, and eq. (25) agrees with
the previous estimate in eq. (23).  
Eq. (25) of course must also be wrong because it gives an
$S$--matrix which is much too small.  The problem with (25) is the same
as with (23): rare, rather than typical, configurations of the
wavefunction dominate the evaluation of the $S$--matrix in center of
mass scattering when $Y>Y_0.$

\subsection{Rare vs typical fluctuations in the Kovchegov equation}

 In order to better appreciate the fact that the role played
in the scattering by the various parts of the wavefunction depends on
the choice of frame we note that, even if we restrict ourselves to
typical configurations, as in Secs. 3.2 and 3.3,
the final result for the $S$--matrix can be
very different if the calculation is performed in a different frame.
For instance, let us choose the asymmetric frame previously used in
Sec. 2, in which one of the participants in the collision is an
elementary dipole while the other participant carries most of the
total rapidity, and thus is highly evolved.  If in this frame we
compute $S_Y$ as the scattering between the elementary dipole and the
typical configuration in the energetic projectile, which is a
condensate, one obtains \cite{ncu} the same result as from the
Kovchegov equation, namely Eq. (14).  (Indeed, one can view Eqs. (7) and 
(8) as imposing the condensate condition on the wavefunction of the high
energy projectile in our discussion in Sec. 2.)  Although this result
is not quite right, as we shall see in the next sections, it is
nevertheless much larger (and thus closer to the correct result) than
the result obtained by working in the center of mass frame, Eq. (25).

 This discussion helps explain why the Kovchegov equation
provides a qualitatively correct result although its derivation
focuses on typical configurations, as we have seen in Sec. 2. 
But this also shows that if one tries to
interpret the result of the Kovchegov equation in a different frame,
like the center of mass frame, this interpretation will generally
correspond to some rare configurations, although not necessarily the
optimal ones.

 Besides the ``dipole frame'' discussed above and in Sec. 2,
where the relevant configurations are those of the condensate, there is
another frame in which the configurations pertinent to
the Kovchegov equation are easily
identified.  For two parent dipoles of the same size $r_0,$ this is
the frame in which the rapidity is equal to $Y-Y_0$ for the
right-mover and to $-Y_0$ for the left-mover.  We shall assume that
the left-mover is subjected to normal evolution, as described by the
JIMWLK equation, in the whole rapidity interval from $-Y_0$ to $0.$
Thus, at the time of scattering, this is a Color Glass Condensate with
saturation scale $Q_s(Y_0) = 1/r_0.$ If $Y=Y_0,$ i.e., if the
right-mover were an elementary dipole, we would have an $S$--matrix of
order one: $S_{Y_0}(r_0) \sim 1.$ The configuration which, for $Y >
Y_0,$ gives the result of the Kovchegov equation is the one in which
the evolution of the right-mover is suppressed in such a way that, in
the whole rapidity range $0<y<Y-Y_0,$ the corresponding wavefunction
consists only of the parent dipole of size $r_0.$ This is a rare
configuration, but has the advantage to involve only one dipole, and
thus give a rather large contribution to $S,$ of order one.  Even
though suppressed by the small probability of the particular
configuration, that we shall shortly compute, this contribution
remains substantially larger than that of a typical configuration,
with $N(Y-Y_0) \gg 1$ gluons (cf. Eq. (24)), which is extremely small (of
the same order as shown in Eq. (25)).

Let $A(r_0, Y-Y_0)$ denote the probability of the rare configuration of
interest.  This is the same as the survival probability of the parent
dipole after a BFKL evolution over a rapidity interval $Y-Y_0$,
and can be computed with the methods of Refs. \cite{uel,ala,lle,tel,Mue,IM03}.
This probability decreases with increasing $Y$ because of gluon radiation,
and the corresponding rate is the same as the virtual term in Eq. (6):
\begin{equation}
{\partial\over \partial y} \,A(r_0,y) \,= \,- \bar{\alpha}_s\int
{d^2z_\perp\over 2\pi}\,{r_0^2\over
(x_\perp-z_\perp)^2(z_\perp-y_\perp)^2}\, A(r_0,y)
\end{equation}
 with $x_\perp$ and $y_\perp$ labelling the quark and
antiquark parts of the dipole $r_0.$ This equation should be
integrated from $y=0$ up to $y=Y-Y_0.$ Note however that the integral
is ill-defined because of poles in the integrand at $z_\perp=x_\perp$
and $z_\perp=y_\perp.$ These divergences reflect the fact that one
cannot avoid the emission of dipoles of arbitrarily small size.

 The situation is similar to that encountered in the discussion
of the Kovchegov equation in the regime where the $S$--matrix is small
(cf. Sec. 2).  Like in that case a physical cutoff, which depends on
$y,$ exists also for the problem at hand.  Let us introduce this
cutoff as a minimal dipole size $\rho(y);$ that is, the integral in
(26) should be restricted to $(x_\perp-z_\perp)^2>\rho^2(y)$ and
$(z_\perp-y_\perp)^2>\rho^2(y)$. Physically this means that, at
rapidity $y,$ we suppress the radiation of dipoles with sizes larger
than $\rho(y)$, but smaller dipoles are allowed\footnote{Note indeed
that the dipoles which are included in the integral in eq. (26) are
those whose radiation is forbidden.}.  Each of these small
radiated dipoles becomes the seed of a normal BFKL evolution going
from the intermediate rapidity $y$ (at which the dipole was emitted)
down to $y=0.$ The evolution is such that new dipoles are produced,
and the typical size of such ``children'' dipoles is increasing with
decreasing $y.$ This is so because the line $\ln(r^2(y)/r_0^2)= -
c\bar{\alpha}_s y,$ where $c$ is given in (11), is a line of constant
amplitude for BFKL evolution \cite{IIM02,MT02}.

\begin{figure}
  \centerline{
  \epsfsize=0.7\textwidth
\epsfbox{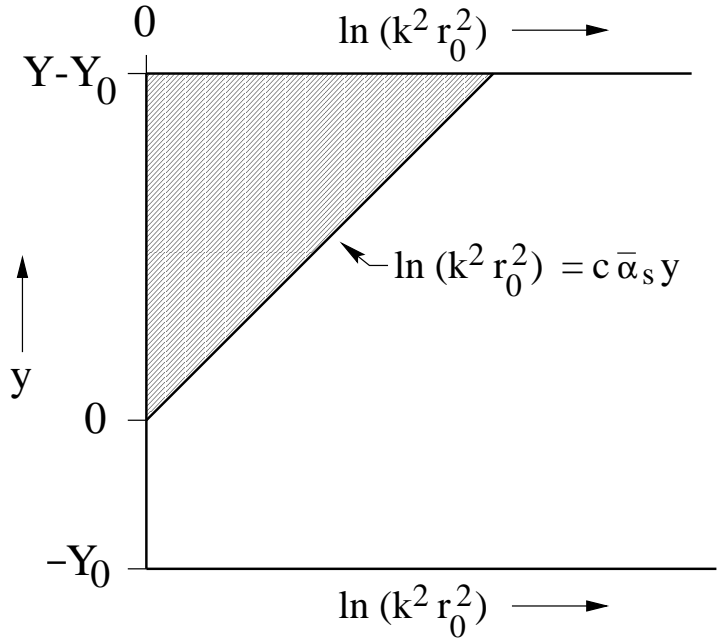}
  }
 \caption[]{The configuration retained by the Kovchegov equation 
in the frame in which the left mover has rapidity $-Y_0$.
}
\end{figure}
We see that, instead of being truly a single dipole configuration,
the configuration that we are constructing also contains small
dipoles, of size $\lesssim \rho(y)$ at the ``moment'' $y$ of their
emission, which then grow up with further decreasing $y,$ and can
become as large as $\rho^2(0) = \rho^2(y) {\rm e}^{c\bar{\alpha}_sy}$ at the
collision ``time'' $y=0.$ Still, this complicated configuration
behaves, during scattering, as the idealized single dipole
configuration, provided all the additional dipoles which are present
in the wavefunction at $y=0,$ and hence also in the interval $0\leq y
\leq Y-Y_0,$ are much smaller than the parent dipole $r_0.$
(Indeed, such small dipoles undergo essentually no scattering, so the
overall $S$--matrix for this configuration
remains very close to $S_{Y_0}(r_0) \sim 1$.) The condition that
$\rho(0)\simle r_0$
implies an upper limit on $\rho(y),$ namely $\rho^2(y)\simle
r_0^2\,{\rm e}^{-c\bar{\alpha}_sy},$ which can also be written as
\begin{equation}
\ln(\rho^2(y)/r_0^2) \simeq - c\bar{\alpha}_s y,
\end{equation}
 since the integral on the right-hand side of (26) is
dominated by the logarithmic regions of integration
where $\rho(y) \ll r \ll r_0$. Here, $r$ is the transverse size of any
of the emitted dipoles, that is, either $|x_\perp-z_\perp|$,
or $|y_\perp-z_\perp|$. 

Thus, the dominant contribution to eq. (26) can be evaluated as:
\begin{equation}
{\partial\over \partial y}\, \ln A(r_0,y) \,\simeq \,-
\bar{\alpha}_s\int_{\rho^2(y)}^{r_0^2}{dr^2\over r^2}\, =\,
\bar{\alpha}_s\ln(\rho^2(y)/r_0^2)\, = - c\bar{\alpha}_s^2y,
\end{equation}
 which after integration over $y$ yields:
\begin{equation}
A(r_0, Y-Y_0)\, \simeq \,{\rm e}^{-{c\over 2}\,\bar{\alpha}_s^2(Y-Y_0)^2}.
\end{equation}
Of course, the evolution towards larger dipoles, with sizes $r \gg r_0$, 
must be forbidden as well. But such large
dipoles give only a small contribution to the integral in eq. (26),
because of the rapid fall off of the integrand at large transverse separations
$|z_\perp - b_\perp| \gg r_0$ (with $b_\perp=(x_\perp + y_\perp)/ 2$).
Specifically, if we denote this contribution as $A'$ (this is an additional
factor which multiplies the previous contribution in eq. (29)), then
$A'$ is estimated as:
$${\partial\over \partial y}\, \ln A'(r_0,y) \,\simeq \,-
\frac{\bar{\alpha}_s}{2}\,r_0^2\int_{r_0^2}{dr^2\over r^4}\, =\,-
\frac{\bar{\alpha}_s}{2}\,.$$
After integration over $y$, this gives a new contribution, 
 of order $\bar{\alpha}_s(Y-Y_0)$, to the exponent of eq. (29).
This contribution is subleading in the regime of interest here, and,
in any case, it goes beyond the general accuracy of the present calculation.
Because of that, dipoles with sizes  $r \ge r_0$ will be ignored in
what follows.

 The $S-$matrix element associated with the particular
configuration of interest
is finally computed as $S_Y=A(r_0, Y-Y_0) S_{Y_0}$ and
coincides, as anticipated, with the result of the Kovchegov equation,
Eq. (14).

  In the next subsection and in Sec. 4 we shall see
that a larger $S-$matrix can be obtained by choosing different rare
configurations, which are also frame dependent.  To that end, it will
be useful to have a graphical representation of the regions of
evolution in the $\big(\!\ln(k_\perp^2r_0^2),Y\big)$ plane, with $k_\perp$
being the momentum conjugate to $r_\perp.$ For the particular
configuration that we have discussed above,
the corresponding plot is displayed in Fig. 1. 
The shaded triangle in this plot represents the
kinematical domain into which dipole emission is forbidden,
and which gives the dominant contribution to $A$, eq. (29).
Note that, up to a factor $\bar{\alpha}_s$, the exponent in
eq. (29) is the same as the area of this shaded triangle.

\subsection{The dominant configuration in the center of mass frame}

  We now turn to the task of finding those
configurations that dominate very high energy dipole-dipole scattering
in the CM frame.  Since the typical configurations that we have previously
considered, in Secs. 3.2 and 3.3, have given an $S$--matrix which is too small
(cf. eqs. (23) and (25)), it is natural to search for
configurations which are more rare in the wavefunction but which lead
to a larger $S$--matrix. The problem with the previous calculations is that the
typical configurations contain too many gluons at the time of collision, 
thus giving a contribution to the $S$--matrix which is extremely small.
This suggests that the dominant configurations for evaluating $S$
at very high energies
should contain less than the mean number of gluons \cite{uel}. 

In view of the discussion in Sec. 3.4, the general strategy for finding the
optimal configurations should be rather clear: Loosely speaking, 
the final goal should be to {\it minimize
the number of gluons} by suppresing the evolution, but at the same time
{\it maximize the probability} of the resulting configuration. The latter
is largest (i.e., of order one) for the configurations subjected to normal 
evolution, but is rapidly decreasing for the configurations in which the 
evolution has been suppressed. Thus, the best results should be obtained
by enforcing the {\it minimal suppression} which still give rise to
a $S$--matrix of order one.

Unfortunately, since we are not able to analyze all the possible configurations,
we cannot transform these considerations into a precise mathematical
criterion for the search of the optimal configurations. What we can do,
however, is to use the experience with the previous example in Sec. 3.4
in order to guess what should be the dominant configurations.
With these configurations, we shall then evaluate the $S$--matrix,
and find a result which is indeed larger than in eq. (14).
Finally, in later sections,
we shall give arguments as to why we believe we have found
the most important wavefunction configurations for evaluating $S$ 
in the high energy limit.

Consider a zero impact parameter collision in the center--of--mass frame at
rapidity $Y > Y_0$. (As usual, $Y_0$ denotes the critical value
for the onset of unitarity corrections, cf. eq. (21).)
For this problem, we shall require that the wavefunction of the
right-moving dipole consist only of the parent dipole, of size $r_0$,
in the rapidity interval $Y_0/2 < y < Y/2$, with a similar requirement
on the wavefunction of the left-moving dipole in the interval
$-Y/2<y<-Y_0/2$. In the rapidity intervals $-Y_0/2<y<0$ for the
left-moving system, and, respectively, $0 < y < Y_0/2$ for the right-moving 
one, we allow normal, BFKL, evolution of the wavefunctions.  
\begin{figure}
\centerline{\epsfig{file=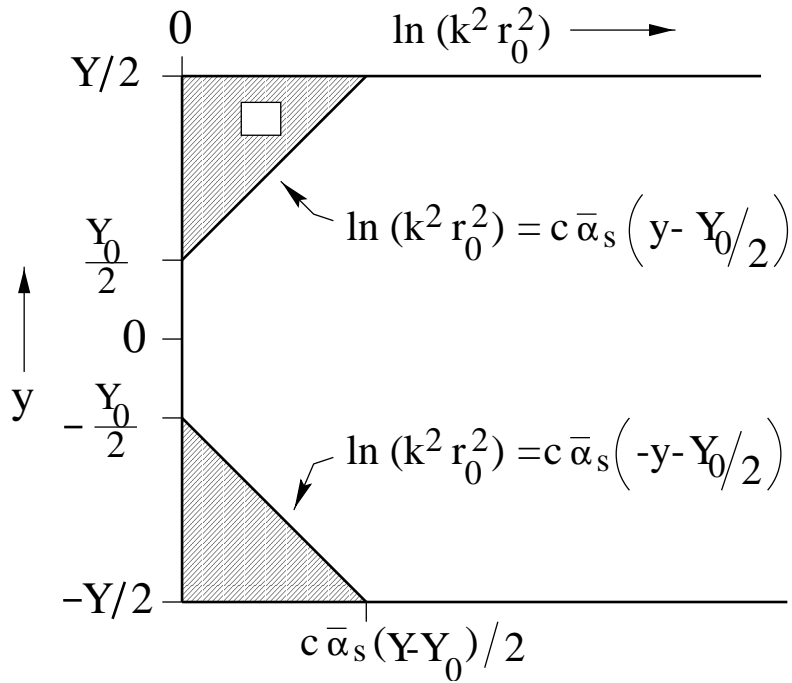,height=9.cm,clip=}}
\caption{The  optimal configuration in the center--of--mass frame.}
\end{figure}

Of course, say in the
$Y_0/2 < y <Y/2$ interval for the right-moving system, we cannot
require that {\it all} evolution in rapidity be absent.  What is required is
that evolution which does exist create only very small dipoles, close to
either the quark or antiquark of the parent dipole, so that in the
interval $Y_0/2 < y < Y/2$ the system have no more than one dipole of
size $\lambda r_0$ or larger, with $\lambda$ a constant of order 1.  In
order to guarantee that this be the case, it is necessary to suppress
the creation of dipoles much smaller than $r_0$ at rapidities $y>Y_0/2$;
otherwise, the dipoles emitted at intermediate rapidities could evolve into 
dipoles of size $r_0$, or larger, at rapidity $Y_0/2$. The
constraints are similar to those we have just discussed in the
previous subsection and are illustrated in Fig. 2.  Gluon emission from
the two parent dipoles, as part of the evolution which forms the left
and right-moving states which scatter on each other, is forbidden if
the gluon has $k_\perp$ and $y$ values lying in the shaded 
triangles\footnote{As discussed in Sec. 3.4, the evolution towards
larger dipoles, with sizes $r \ge r_0$ (or $\ln (k_\perp^2 r_0^2) < 0$), 
is also forbidden, but this can be neglected in the calculation of the
probability $A(r_0, {(Y-Y_0)/ 2})$.}
of Fig. 2.  (For the moment ignore the small unshaded square in the
upper triangle.)  The line
\begin{equation}
\ln (k_\perp^2 r_0^2) = c\bar{\alpha}(y-Y_0/2),
\end{equation}
 and a similar line for the lower triangle, is determined, as
we have just seen, by the requirement that gluons lying to the right
of that line, for any $y > Y_0/2,$ cannot evolve through normal BFKL
evolution to give gluons having $k_\perp \lesssim 1/r_0$ at rapidity
$Y_0/2.$ The constant $c$ is again the same as in (10) and so the line
(30) is a line of constant amplitude in terms of BFKL evolution. Thus
evolution which crosses this line has a small probability.

 Still as in Sec. 3.4, we shall denote by $A(r_0, {(Y-Y_0)/ 2})$ the
probability that the parent dipole, of size $r_0$ and rapidity
$Y/2$, not give rise to any emissions in the upper triangle of Fig. 2.   
The $S$--matrix is then
given by a factor $A(r_0, {(Y-Y_0)/ 2})$ for each of the parent
dipoles partaking in the collision, times the (partial) $S$--matrix 
for the scattering of two dipoles separated by a rapidity gap
$Y_0/2 - (-Y_0/2)=Y_0$, and which are subjected both to normal evolution.
The latter is, of course, $S_{Y_0}(r_0)\sim 1$. After also using
eq. (29) for $A(r_0, {(Y-Y_0)/ 2})$, one finally obtains:
\begin{equation}
S_{Y}(r_0)\simeq {\rm e}^{-{c\over 4}\,\bar{\alpha}_s^2(Y-Y_0)^2}
\,S_{Y_0}(r_0)
\end{equation}
a result which is much larger than (25), and also significantly
larger than the
result (14) coming from the Kovchegov equation. 

Thus, one can avoid a very small $S-$ matrix, as given in
(25), by not forgetting the rare parts of the wavefunction which can
dominate the $S-$matrix when $S$ is small. These rare configurations
consist of unusually small numbers of gluons being present in the
wavefunction.  

It should be emphasized that (31) holds only when the
parent dipoles have a fixed size, $r_0$ in our case.  For instance,
if one were to scatter heavy onia states, eq. (31) would not emerge.
In the case of heavy onia there is a wavefunction giving the probability of
having a parent dipole of a particular size, $r$. Because of that, the
$S$--matrix for the scattering of
two onia at very high energies would be dominated by
parent dipole sizes $r$ of order $1/Q_s(Y)$.

\subsection{Sampling other configurations}

 In the last section we found a particular configuration
that leads to an $S-$matrix which is much larger than that given by
the Kovchegov equation.  While this is sufficient to prove that
fluctuations must be important, and thus that the result in eq.
(4) cannot be exact, it
is, perhaps, not clear that we have found the optimal configurations
for dipole-dipole scattering in the CM frame.  In this section we give
arguments to the end that (31) is the correct answer and that the
configurations we have focused on are indeed the dominant ones.

The first change one may consider in the previous calculation would
be to modify the values of the intermediate rapidities $\pm Y_0/2$
which separate between suppressed and normal evolution.
So, let us replace $Y_0 \to Y_1$, with $Y_1\ne Y_0$, in Fig. 2 and
the related calculations. There are two possibilities --- $Y_1 < Y_0$,
or $Y_1 > Y_0$ ---, but as we argue now they both lead to a $S$--matrix
which is much smaller than that in eq. (31). {\it i\,}) If $Y_1 < Y_0$,  
the probability $A(r_0, {(Y-Y_1)/ 2})$ for the suppressed configuration
is even smaller than before, while the partial $S$--matrix $S_{Y_1}$ for the
collision between two normally evolved systems is still of order one
(since $Y_1 < Y_0$ corresponds to weak scattering). Thus, this situation
is clearly less favourable. {\it ii\,}) If $Y_1 > Y_0$, the probability
$A(r_0, {(Y-Y_1)/ 2})$ is larger, but this gain is more than compensated by
the strong decrease in the partial $S$--matrix $S_{Y_1}$. To see this, note
that for the calculation of $S_{Y_1}$ we are either in the situation
described in Sec. 3.2 (if $Y_1 < 2Y_0$), 
or in that of Sec. 3.3 (if $Y_1 > 2Y_0$), and, as we have seen, both
these situations give only tiny contributions to $S$.
If we focus on $|\ln S_Y|$, for definiteness, then $|\ln S_{Y_1}|$
grows exponentially with the difference $Y_1- Y_0$ (cf. eqs. (22) or (25)), 
while the corresponding decrease in  $|\ln A|$ is only quadratic.
Clearly, the overall $S$--matrix decreases rapidly with increasing
$Y_1- Y_0$.

\begin{figure}
\centerline{\epsfig{file=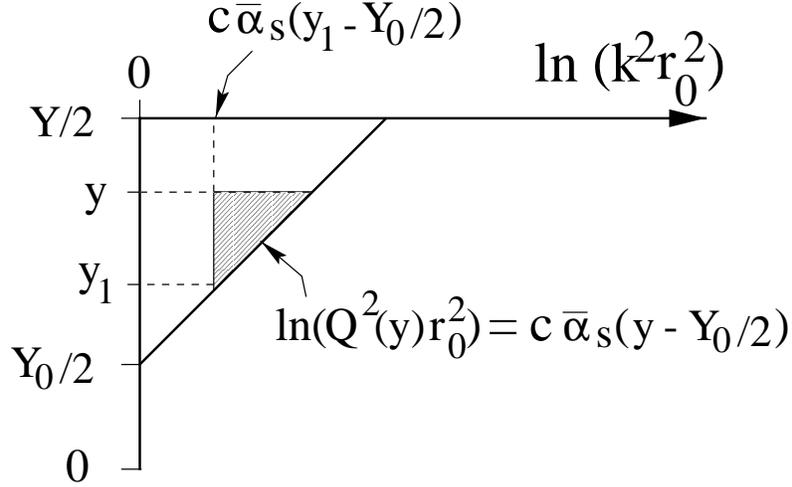,height=6.5cm,clip=}}
\caption{The shaded area is the domain of suppressed evolution
for a parent dipole within the small unshaded rectangle in Fig. 2.}
\end{figure}
  The other major change that we could make in our 
calculation in the last section would be to allow some limited
emission of gluons into the forbidden triangles of Fig. 2.  
Recall that the  area of such a triangle determines the
penalty factor $A$ associated with suppressing the evolution
(cf. the remark at the end of Sec. 3.4). Then let us
calculate the change in our previous result which occurs if we allow
emission into the small square in the upper triangle of Fig. 2.  
Take the area of the square to be $1/\bar{\alpha}_s$, so that now one gets
a factor $A$ which is a factor of $e$ larger than the previous one:
\beq
A^\prime\,= \,{\rm e}\times A(r_0, {(Y-Y_0)/ 2}),\eeq
with $A(r_0, {Y-Y_0\over 2})$
given by eq. (29). We have less suppression than before because 
we have {\it not} forbidden the parent dipole to emit into the 
small square.  However, now we must forbid the newly
created (small) dipole from evolving further and again creating a
strong suppression like that given in (25).  In Fig. 3, the shaded
triangle represents the region into which one must forbid the
additional dipole from emitting gluons\footnote{Once again, we consider only
that forbidden area which gives the dominant contribution to the probability $A$; 
see eq. (35).}. If $y$ and $k_\perp$ are the
coordinates of the gluon emitted by the parent dipole
(i.e., the central coordinates of the small
unshaded square in Fig. 2), and if $y_1$ is defined by (see Fig. 3):
\begin{equation}
\ln(k_\perp^2 r_0^2) = c\bar{\alpha}_s(y_1-Y_0/2),
\end{equation}
 then the shaded area in Fig. 3 --- the area into which emission
from the secondary dipole must now be forbidden --- is
\begin{equation}
a\, =\, {1\over 2}\, c\bar{\alpha}_s(y-y_1)^2\,.
\end{equation}
The probability of not emitting into this area is:
\begin{equation}
A'' \,=\, {\rm e}^{-\bar{\alpha}_sa}\,=\, 
{\rm e}^{-{c\over 2}\,\bar{\alpha}_s^2(y-y_1)^2}.
\end{equation}
In order for this suppression not to be 
larger than the gain from being allowed to have the secondary emission,
cf. eq. (32),
one must require that $\bar{\alpha}_s a < 1$. This
implies the following condition on the location of the small square in
Fig. 2 :
\begin{equation}
\Big[\ln\big(Q^2(y)r_0^2\big) - \ln \big(k_\perp^2r_0^2\big)\Big]^2
\,< \,2c,
\end{equation}
where, for more clarity, we have denoted the rightmost borderline of the suppressed
area in Fig. 3 as (cf. eq. (30)): 
$\ln \big(Q^2(y) r_0^2\big) = c\bar{\alpha}(y-Y_0/2)$.

Eq. (36) shows that, for a given $y$, the point
$\ln (k_\perp^2r_0^2)$ lies within
one unit of the boundary, $\ln \big(Q^2(y)r_0^2\big)$.
But since the boundary is ambiguous at this level of precision 
--- e.g., the right hand side of eq. (30) is
specified only up to corrective terms of order one ---,
we see that it does not seem possible to relax
the assumption we made in deriving eq. (31), namely that there must be
no emission into the triangle regions of Fig. 2.

\section{The picture in a general frame}

  In this section we show how the result (31) comes
about in an arbitrary frame. In order to describe the frame choice
and the scattering picture, it is useful to refer to Fig. 4.  We scatter
a left-moving parent dipole of size $r_1$ and rapidity $-Y_2$ on
a right-moving parent dipole having size $r_0$ and rapidity $Y-Y_2$.
Recall that $Y_0$ is the value of the rapidity difference between 
two such dipoles
where the $S$--matrix begins to be significantly different from one.
We also suppose that $Y_2 \le {1\over 2}(Y-Y_0)$, for later convenience.
\begin{figure}
\centerline{\epsfig{file=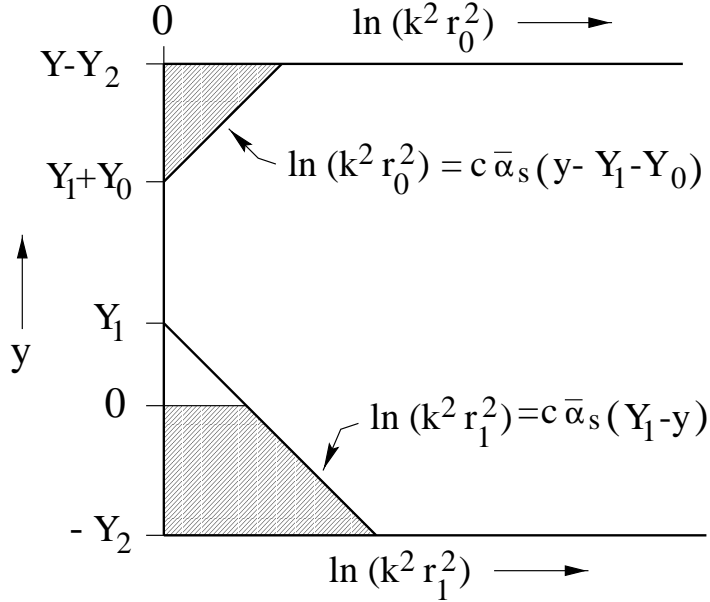,height=8.cm,clip=}}
\caption{Choosing the  optimal configuration in some generic frame.}
\end{figure}
 
Clearly one cannot allow both dipoles $r_0$ and $r_1$ to
have any significant amount of normal BFKL evolution,  
or else we will find a
strong suppression of the type given by (25). Since the left-moving
dipole has the smaller rapidity it is natural (and easier) to suppress
its evolution, and we will describe that suppression in a moment.  As
for the dipole $r_0$, we suppress evolution over its $Y-Y_2-(Y_1+Y_0)$
highest units of rapidity exactly as we did in Sec. 3, with the region
of suppression given by the upper shaded triangle of Fig. 4.  In the
lowest $Y_0+Y_1$ units of rapidity the parent dipole undergoes normal
evolution.  We will determine $Y_1$ later, by maximizing the $S$--matrix. 
The unshaded
triangle, whose rapidity values go from $0$ to $Y_1$, denotes the
saturation region for the right-mover,
that is, the region where the parent dipole $r_0$
has evolved into a Color Glass Condensate (so the evolution is non-linear
within that particular region). Let $Q_s\equiv Q_s(Y_0+Y_1)$ be the saturation 
scale which characterizes this condensate. Since obtained after a
normal evolution through $Y_0+Y_1$ units of rapidity, this is of the form:
\beq
Q_s^2(Y_0+Y_1)\,=\,Q_0^2\,{\rm e}^{c\bar{\alpha}(Y_0+Y_1)},\eeq
where $Q_0^2$ is an intrinsic scale of the right-moving system, proportional 
to $1/r_0^2$ and, possibly, powers of $\alpha_s$. By the very definition
of $Y_0$, we have $Q_s^2(Y_0) = 1/r_1^2$, so the saturation momentum
(37) can be rewritten as:
\beq
Q_s^2\,\equiv\,Q_s^2(Y_0+Y_1)\,=\,{1\over r_1^2} \,{\rm e}^{c\bar{\alpha}Y_1}\,,\eeq
where the dependence upon the unknown scale $Q_0^2$ has disappeared.

Turning now to the parent dipole $r_1$,
 we require that no additional dipoles be created, through
gluon emission, which would have a strong interaction with the
right-moving Color Glass Condensate. Clearly, the dipoles 
which would have such strong interactions are those which,
at the time of scattering (i.e., at $y=0$), would be of size 
$r\simge 1/Q_s$.
This means that, at all the intermediate rapidities within the range
$-Y_2 < y < 0$, we have to suppress the emission of those dipoles which,
after a normal evolution over the last $|y|$ units of rapidities, could
become of size $1/Q_s$, or larger. By referring to the similar discussion
in Sec. 3.4, we see that the maximal size $\rho(y)$ allowed for
a dipole emitted at rapidity $y$ is constrained by
\beq
\rho^2(y)\,\simle\, \frac{1}{Q_s^2}\, {\rm e}^{-c\bar{\alpha}_s|y|}\,=\,
r_1^2\, {\rm e}^{-c\bar{\alpha}_s(Y_1 -y)},\eeq
where we have also used eq. (38) together
with the fact that  $|y|=-y$. In terms of the conjugate momentum variable
$k_\perp$, this gives the lower shaded
region in Fig. 4 as the region into which radiation is forbidden. 

We are now in a position to estimate the $S$--matrix as
\begin{equation}
{S}_Y(r_0,r_1)\,=\,A_R(r_0,Y-Y_0-Y_1-Y_2) \,{\cal S}_{Y_0+Y_1}(r_0,r_1)\,
A_L(r_1,Y_2)\,,
\end{equation}
 where $A_R$ and $A_L$ are the suppression factors from the
no emission requirements for the two dipoles, and
 ${\cal S}_{Y_0+Y_1}(r_0,r_1)$ is the
$S-$matrix for scattering an elementary dipole of size $r_1$ on a
dipole $r_0$ which has evolved into a Color Glass Condensate
characterized by the saturation momentum (38).

To compute ${\cal S}$, one can rely on the Kovchegov equation,
or, more exactly, on its approximate form in eq. (9),
valid deeply at saturation. This is correct for the present purpose since,
as discussed at the beginning of Sec. 3.4, the  Kovchegov equation describes
correctly the scattering between an elementary dipole (here, the dipole
$r_1$) and the typical configurations in a Color Glass Condensate
(here, the dipole $r_0$ which has been allowed to carry out its normal 
evolution over the rapidity interval $Y_0+Y_1$). 
By using eq. (14), one obtains: 
\begin{equation}
{\cal S}_{Y_0+Y_1}(r_0,r_1)\,\simeq\,{\rm e}^{-{c\over 2}\,\bar{\alpha}_s^2Y_1^2}
\,S_{Y_0}(r_0)\,.
\end{equation}
We note in passing that the exponent in
(38) is, except for a factor of $\bar{\alpha}_s,$ the area of the
unshaded triangle in Fig. 4, and where, as usual, the exponents of
$A_R$ and $A_L$ are given in terms of the area of the upper and lower
shaded regions of Fig. 4.  Thus
\begin{equation}
A_R\,=\,{\rm e}^{-{c\over 2}\,\bar{\alpha}_s^2(Y-Y_2-Y_1-Y_0)^2},
\end{equation}
 and, similarly,
\begin{equation}
A_L\,=\,{\rm e}^{-{c\over 2}\,\bar{\alpha}_s^2[(Y_1+Y_2)^2-Y_1^2]}.
\end{equation}
By combining eqs. (40)--(43), one obtains:
\begin{equation}
S_Y\,=\,\exp\left
\{-{c\over 2}\,\bar{\alpha}_s^2\,\big[(Y-Y_2-Y_1-Y_0)^2+(Y_1+Y_2)^2\big]\right\}
S_{Y_0}(r_0)\,.\end{equation}
 
The $S$--matrix given by (44) refers to a particular set of
configurations of the wavefunctions, characterized by one parameter:
the rapidity $Y_1$ which describes the
amount of evolution in the right-moving system. (Recall that
the other rapidity parameter in eq. (44), namely $Y_2$, specifies
the frame of reference.) The actual $S$--matrix
should be determined by the value of $Y_1$ which maximizes the right
hand side of eq. (44), or, equivalently, which minimizes the exponent
there. This condition yields:
\begin{equation}
Y_1\,=\,{1\over 2} \,(Y-Y_0) - Y_2.
\end{equation}
This is non-negative, because of our initial assumption
$Y_2 \le {1\over 2}(Y-Y_0)$.

Finally, $S_Y$ is obtained by evaluating  eq. (44)
with the optimal value for  $Y_1$, eq. (45) :
\begin{equation}
S_Y(r_0,r_1)\,=\,{\rm e}^{-\frac{c}{4}\,\bar{\alpha}_s^2(Y-Y_0)^2}\,S_{Y_0}(r_0).
\end{equation}
(The right hand side depends upon $r_1$ via the ``critical'' rapidity $Y_0$.)
This result is independent of $Y_2$ (i.e., upon the choice of a frame), 
and is exactly the same as the 
corresponding result in the CM frame, eq. (31).

In fact, the picture of the evolution for the optimal configuration
in the CM frame, as shown previously in Fig. 2, can be obtained from
the general picture in Fig. 4 by first choosing $Y_2$
in such a way that $Y_1=0$ --- 
this requires $Y_2=(Y-Y_0)/2$ (cf. eq. (45)) ---, and then
performing a supplimentary boost by a rapidity amount $\Delta Y=-Y_0/2$
(so that $Y_2 \to Y/2$, as in Fig. 2). The latter operation changes 
only the position of 0 on the rapidity axis, but not also the 
picture of the evolution\footnote{Indeed, if $Y_1=0$, there is no
non-linear evolution involved, but only linear, BFKL, evolution, which
is invariant under a small boost.}.

In particular, it is instructive to visualize the evolution in the
``dipole frame'', i.e., the rest frame of the left-mover ($Y_2=0$),
in which both the optimal configuration (cf. Fig. 4) and the configuration
retained by the Kovchegov equation are explicitly known.
These configurations are illustrated in  Figs. 5.a and b, respectively. 
As usual, a shaded area represents a kinematical domain into which dipole 
emission is forbidden,
while an unshaded triangle represents the saturation region.
Also, up to a factor of $\bar{\alpha}_s,$ the total area of the 
(shaded and unshaded) triangles represents the exponent of the
$S$--matrix associated with the respective configuration.
As manifest on these figures, the area of the empty triangle in Fig. 5.b
is twice as much as the sum of the areas of the two small
triangles in Fig. 5.a. 

Note also that, for both configurations in Fig. 5, the right-mover
ends up as a Color Glass Condensate at $y=0$.
But while the configuration
retained by the Kovchegov equation (cf. Fig. 5.b) corresponds to
normal evolution over the whole rapidity range --- BFKL evolution 
from $y=Y$ down to $y=Y-Y_0$, and then non-linear, JIMWLK, evolution
over the remaining $Y-Y_0$ units of rapidity ---, in the optimal 
configuration (cf Fig. 5.a), the normal evolution is allowed only over
the lower $(Y+Y_0)/2$ units of rapidity.
As a result, the saturation scale which characterizes the
condensate in the optimal configuration, namely, $Q_s((Y+Y_0)/2)$, is much
smaller than the corresponding scale $Q_s(Y)$ for the normal evolution.
\vspace*{.5cm}
\section*{Acknowledgments}

Much of this work was done
while one of the authors (A.M.) was a visitor at LPT (Universit\'e de
Paris XI, Orsay). He wishes to thank Dr. Dominique Schiff for her 
hospitality and support during this visit.

\begin{figure}
\centerline{\epsfig{file=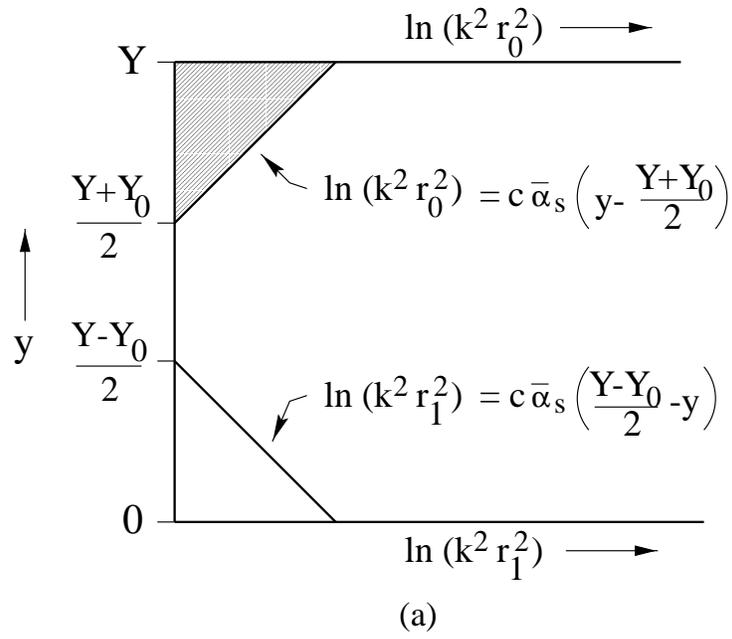,height=8.5cm,clip=}}
\vspace*{1.cm}\centerline{
\epsfig{file=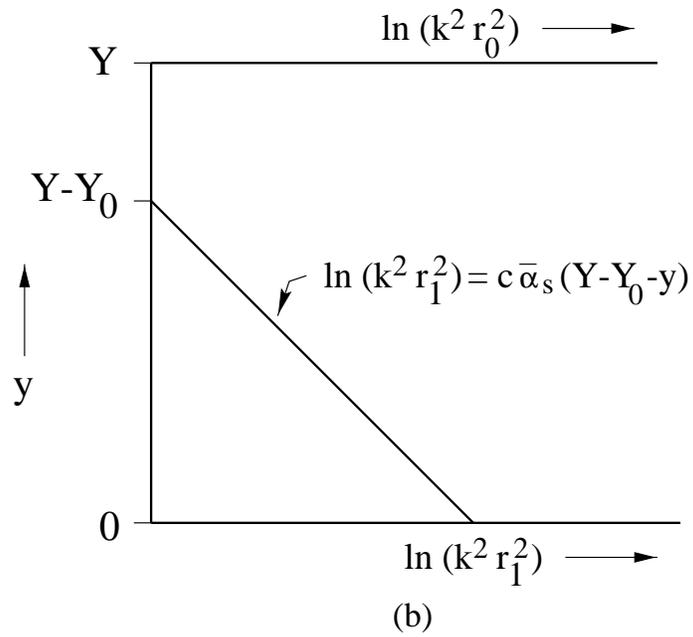,height=8.5cm,clip=}}
\caption{The  optimal configuration (above) and the
configuration  retained by the Kovchegov equation
(below) in the frame in which $Y_2=0$.}
\end{figure}

\end{document}